\documentclass[twocolumn,trackchanges]{aastex701}

\begin{document}

\title{Pre-perihelion Volatile Evolution of Interstellar Comet 3I/ATLAS Indicating Significant Contribution from Extended Source in the Coma}

\author[orcid=0009-0001-7333-5202]{Juncen Li}
\affiliation{Shanghai Astronomical Observatory, Chinese Academy of Sciences, 80 Nandan Road, 200030 Shanghai, People's Republic of China}
\email[show]{jcli@shao.ac.cn}

\author[orcid=0000-0002-4120-7361]{Xian Shi}
\affiliation{Shanghai Astronomical Observatory, Chinese Academy of Sciences, 80 Nandan Road, 200030 Shanghai, People's Republic of China}
\email[show]{shi@shao.ac.cn}

\author[orcid=0000-0001-9067-7477]{Man-To Hui}
\affiliation{Shanghai Astronomical Observatory, Chinese Academy of Sciences, 80 Nandan Road, 200030 Shanghai, People's Republic of China}
\email{mthui@shao.ac.cn}

\author[orcid=0000-0001-5654-5972]{Jianchun Shi}
\affiliation{Shanghai Astronomical Observatory, Chinese Academy of Sciences, 80 Nandan Road, 200030 Shanghai, People's Republic of China}
\email{jcshi@shao.ac.cn}


\begin{abstract}

Interstellar comets provide rare opportunities for probing the diversity of refractory and volatile inventory around other stars. 
As the second ever interstellar comet, and the third interstellar object, 3I/ATLAS has been the focus of telescopic observations since its discovery in July 2025. 
Following the previous observations at multi-wavelengths, we present further radio observations of the 1665/1667 MHz ground-state OH lines and millimeter observations of the CO($J$=1-0) transition at 115.271 GHz that trace the coma $\rm H_2O$ and CO abundances, respectively. 
We derived OH production rates of $(1.32\pm0.47)\times10^{28}\ \rm s^{-1}$ at 2.27 au and $(1.89\pm0.37)\times10^{28}\ \rm s^{-1}$ at 1.96 au as well as an average CO production rate of $\rm (5.75\pm1.91) \times 10^{27}\, s^{-1}$ between 2.33 and 1.75 au, inferring a CO/$\rm H_2O$ ratio of ($28\pm11\%$).
With the mean HCN production rate of $2.5\times 10^{25}\ \rm s^{-1}$ at 2.1 au reported by \citet{2025arXiv251120845R} and \citet{2025arXiv251002817C}, we infer a CO/HCN ratio of ($230\pm76$). 
By synthesizing water production rates measured with instruments of different apertures, we found that the sublimation from extended source in the coma contributes significantly to 3I's pre-perihelion water measurements, accounting for up to 80\% from 3 au to 2 au.

\end{abstract}



\section{Introduction} 
\label{intro}

The discovery of interstellar objects (ISO) traversing the Solar System has opened a new window on the material inventory and planetesimal formation and evolution around other stars. 
While the first ISO 1I/‘Oumuamua showed no detectable coma and left its volatile composition ambiguous \citep{2017Natur.552..378M}, the second 2I/Borisov displayed a ``classical'' cometary coma and tail \citep{2019ApJ...886L..29J}, enabling the first direct compositional measurements of an interstellar comet \citep{2019ApJ...885L...9F,2020ApJ...893L..48X,2021AA...650L..19O,2021NatAs...5..586Y}. 
Observations at different wavelengths of 2I revealed a suite of volatiles and radicals and, notably, relatively high CO abundances compared to many Solar System comets observed at similar heliocentric distances \citep{2020NatAs...4..861C,2020NatAs...4..867B}. 
These measurements suggest that some interstellar comets may have formed in colder regions or experienced prolonged residence at very low temperatures, preserving CO-rich ices.

Discovered on July 1, 2025 by the Asteroid Terrestrial-impact Last Alert System (ATLAS), 3I/ATLAS (C/2025 N1, here after 3I) is the third confirmed ISO after 1I/‘Oumuamua and 2I/Borisov \citep{2025MPEC....N...12D,2025ApJ...989L..36S}.
Its discovery immediately attracted worldwide attention due to its unprecedented orbital properties, including an eccentricity exceeding 6 and a hyperbolic velocity over 50 km/s \citep{2025MNRAS.542L.139B}.
According to the \={O}tautahi-Oxford model, 3I originated from the Milky Way's thick disk, being the first one from that population entering our Solar System \citep{2025ApJ...990L..30H}. 

3I displayed the signs of a dust coma already at its discovery distance of 4.53 au \citep{2025ATel17264....1A}, making it the second identified interstellar comet after 2I/Borisov.
Extensive observations have been carried out since 3I's discovery, spanning ultraviolet to infrared wavelengths, accumulating measurements of its developing activity.
Clear signs of dust activity were detected by Hubble Space Telescope at 3.8 au, suggesting a mass loss rate of 6 to 60 kg/s \citep{2025ApJ...990L...2J}.
However, no detection of C$_{2}$, NH$_{2}$, CN, or [OI] were achieved at 3.46 au \citep{2025MNRAS.544L..31O}. 
Furthermore, spectroscopic observations between 4.4 to 4.0 au suggest 3I's dust particles have a relatively high ice fraction of $\sim$30\%\citep{2025ApJ...992L...9Y} with a redder coma in optical spectrum than most Solar System comets \citep{2025AA...700L..10A,2025ApJ...990L..65K,2025MNRAS.544L..31O}, and a $\rm CO_2$ dominated gas coma was detected by JWST and SPHEREx \citep{2025ApJ...991L..43C,2025RNAAS...9..242L}.
These studies show that 3I is markedly different from its predecessor 2I, or Solar System comets.

To understand the nature of 3I, the three primary species ($\rm H_2O$, CO, and $\rm CO_2$) content makes a key target of investigation as it helps determine its formation conditions and contributes to our understanding of molecular distribution in the original planetary system of this ISO.
Among the three, ground-based monitoring of $\rm H_2O$ and CO vapour in 3I's coma is feasible. 
OH radicals, being the direct photodissociation products of $\rm H_{2}O$ molecules, serve as the primary tracer for water production rates in comets. 
The detections of OH in 3I's gas coma were performed at different wavelengths, with the first detection by the SWIFT space telescope at ultraviolet \citep{2025ApJ...991L..50X} and follow-up at optical and radio wavelength \citep{2025arXiv250926053H,2025ATel17473....1P,2025ATel17499....1P,2025CBET.5625....1C}.
In this work, we report the results of pre-perihelion monitoring of 3I’s CO and $\rm H_2O$ with the detections of CO(1-0) emission lines at 115 GHz and the earliest detection of OH at 1.6 GHz.
Section 2 provides a description of the observations. 
Section 3 presents our results from these data. 
Section 4 discusses the evolution of $\rm H_2O$ production rate and the abundance of CO in 3I/ATLAS, and places this interstellar object into context with solar system comets measured to date.


\section{OBSERVATIONS AND DATA REDUCTION} \label{sec:obs}

The 115 GHz measurements of CO were obtained from mid-August to September 2025 with the 13.7-m Millimeter-Wave Telescope in Delingha, China (see details in \citealt{2020AJ....159..240W,2025AJ....169..126L}), for a total effective integration time of more than 10 hours. 
The HPBW (half power beam width) is $\sim$50$^{\prime \prime}$, while the main beam efficiency $\eta_{B}$ was $\sim$50\% at this frequency of 115 GHz.
Observations of OH at 1.6 GHz were performed using the Tianma 65-m Radio Telescope (TMRT; see \citealt{2025AA...701A.204L}) on 7 individual days in August and September, totaling above 10 hours of effective integration.

The pointing accuracy was verified daily using strong calibrator sources: IRC+10216 for the 115 GHz observations and W49N for the 1.6 GHz observations.
Observation geometries of 3I were obtained using Horizons System of Jet Propulsion Laboratory (JPL)\footnote{https://ssd.jpl.nasa.gov/horizons/} and the Minor Planet Ephemeris Service (MPEC)\footnote{https://minorplanetcenter.net/iau/MPEph/MPEph.html}.
An observing log for all observations of 3I during these periods are shown in Table \ref{tab:1}.

Data reduction and analysis were conducted with the CLASS package, part of the GILDAS software suite\footnote{http://www.iram.fr/IRAMFR/GILDAS}. 
The processing included Doppler correction for the comet’s radial velocity relative to the telescope during the observations (see Table~\ref{tab:1}). 
Considering the strong radio frequency interference in L band during our observations, spectral analysis was performed using only the right circular polarization data sometimes, as the left circular polarization data were significantly affected. 
A linear baseline was subtracted from each spectrum, excluding the region of the emission line. 
Individual spectra were then averaged using weights based on their rms noise levels. 


\begin{deluxetable}{ccc}
\tablewidth{0pt} 
\tablecaption{Characteristic of Target Spectral lines \label{tab:2}}
\tablehead{
\multirow{2}{*}{Species} & \multirow{2}{*}{Transition} & \colhead{Frequency}  \\
\colhead{} & \colhead{} & \colhead{$[\rm GHz]$}   
}
\startdata
\multirow{2}{*}{OH($^2\Pi_{3/2}$)} & \colhead{$J=3/2,\ F=1-1$} & \colhead{1.665} \\
\colhead{} & \colhead{$J=3/2,\ F=2-2$} & \colhead{1.667} \\
$^{12}$CO & $J=1-0$ & 115.271 \\
\enddata
\end{deluxetable}

\section{RESULTS} \label{sec:results}

\begin{figure}
\centering
    \centering
    \includegraphics[width=0.45\textwidth]{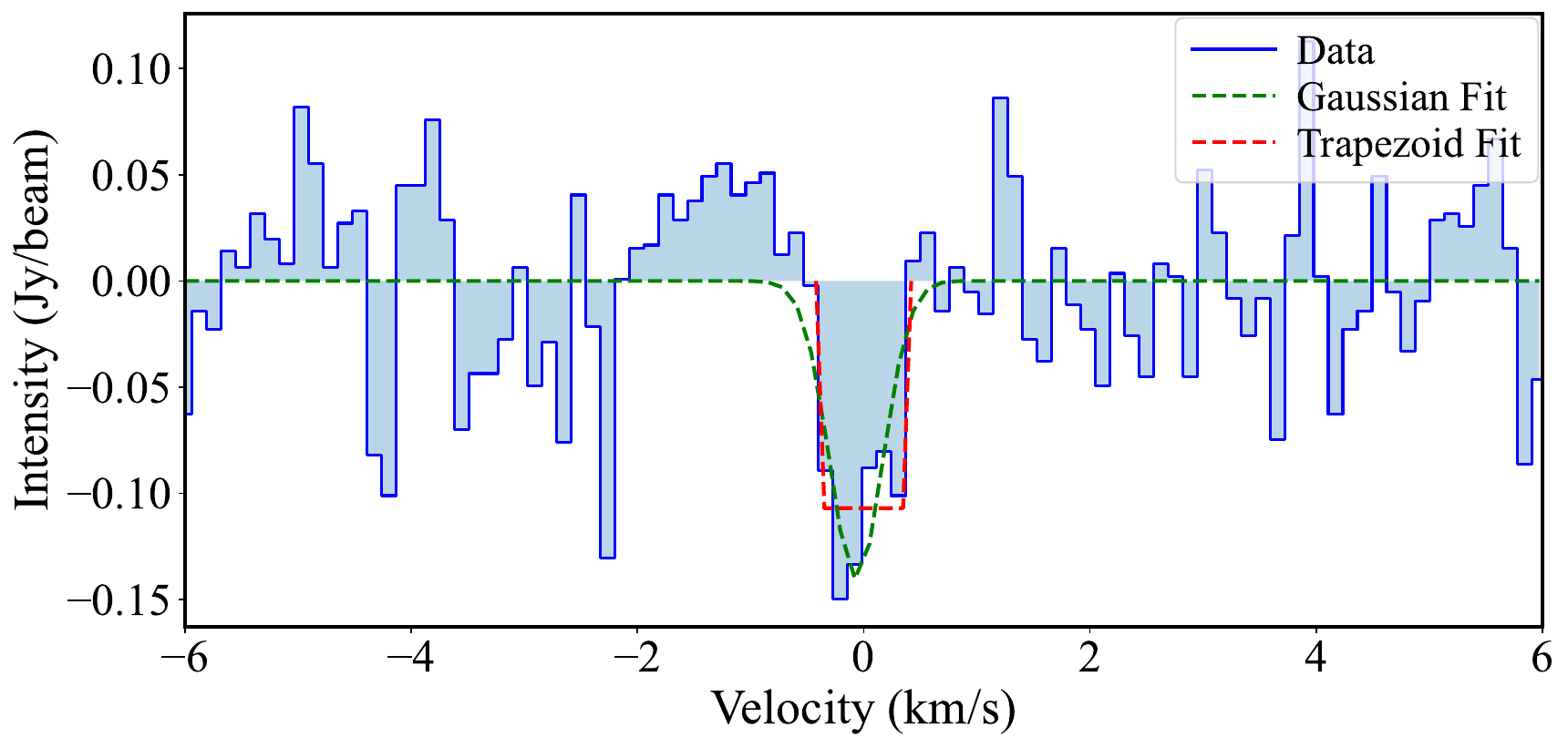}
    \includegraphics[width=0.45\textwidth]{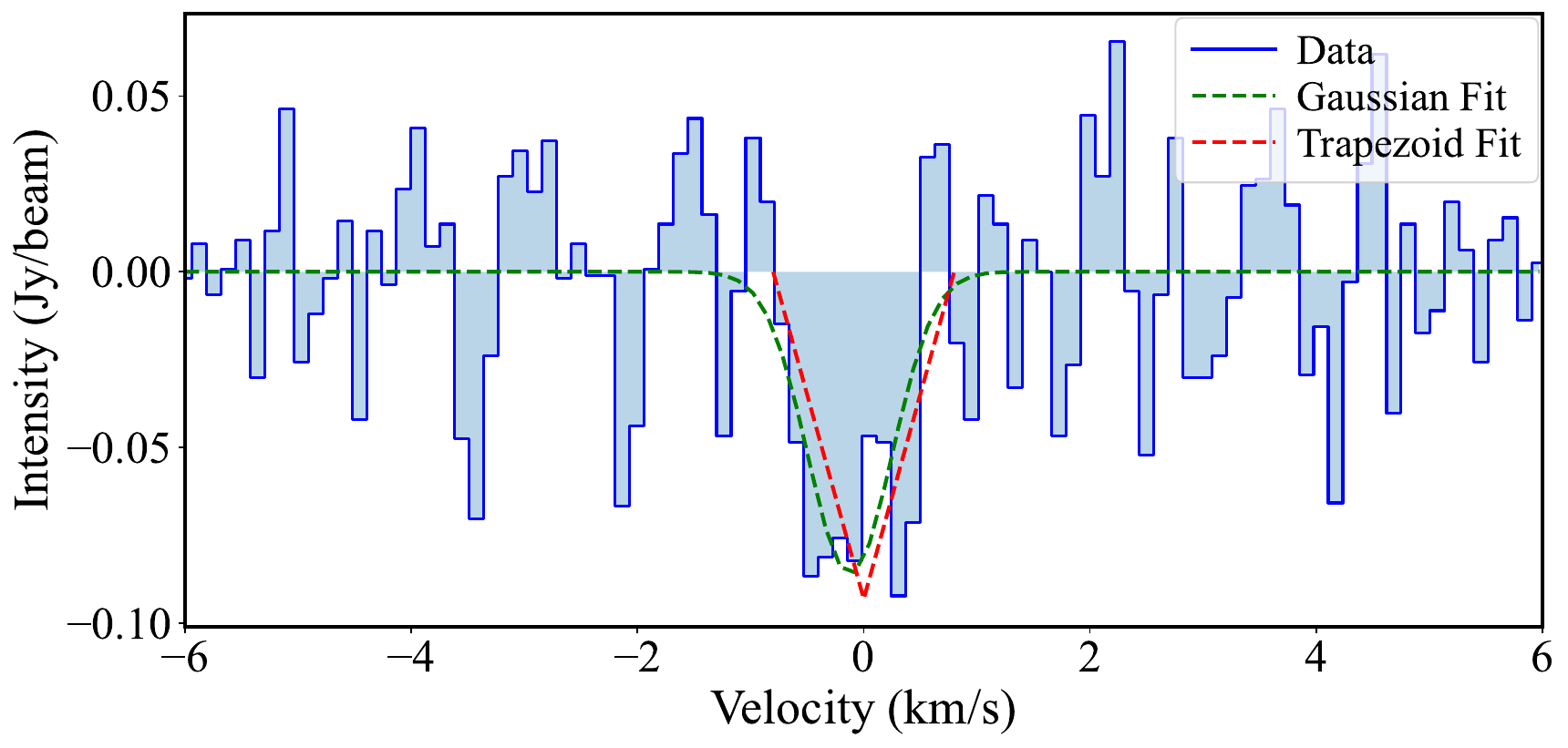}
    \caption{Averaged 18 cm OH lines (scaled to 1667 MHz) of 3I obtained in four observations with both a Gaussian and a trapezoid fit: Sep. 08 and 09 (top), Sep. 18 and 23 (bottom). Original observations are shown in Figure \ref{fig:all2} and \ref{fig:all3}.}
    \label{fig:1}
\end{figure}

\begin{figure}
\centering
    \centering
    \includegraphics[width=0.45\textwidth]{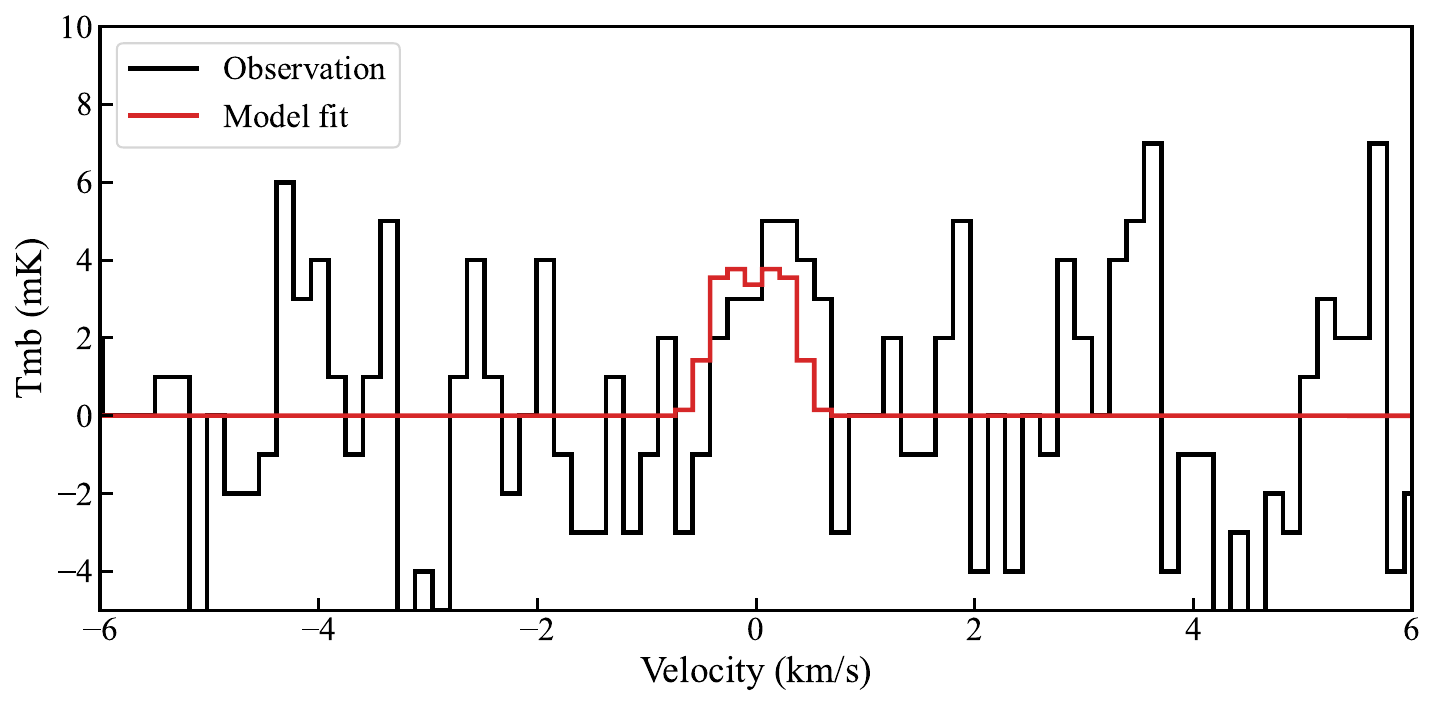}
    \caption{Averaged CO(1-0) line of 3I obtained from September observations. Original observations are shown in Figure~\ref{fig:allco}.}
    \label{fig:2}
\end{figure}

To enhance the overall signal-to-noise ratio (S/N) given the limited daily integration time, spectra were averaged using weights derived from the noise level of each individual day. 
By integrating observations of Sep. 8 and 9, we first detected OH radical lines at 1667 MHz at pre-perihelion heliocentric distances ($r_h$) of around 2.27 au at above $3\sigma$ (Fig.~\ref{fig:1}). 
Subsequently, another detection of OH was obtained at around 1.96 au by integrating signals on Sep. 18 and 23.
The CO(1-0) line was weaker at a level about $3\sigma$ by integrating several days of observations between Sep. 07 and Sep. 29 when 3I was traveling inbound from 2.33 to 1.75 au (Fig.~\ref{fig:2}).

\subsection{OH lines analysis}

The integrated intensity was computed over the velocity interval [$-1,1$] $\rm km \cdot s^{-1}$ with its uncertainty estimated from $\sqrt{n} \times {\rm RMS} \times dv$, where $dv$ is the channel width and $n$ is the number of channels covering the line.
Their values are summarized in Table~\ref{tab:3}.

The OH line profiles were analyzed using the symmetric  standard trapezoid model for cometary radio observations (e.g., \citealt{2002AA...393.1053C,2025AA...701A.204L}). This model derives the OH velocity ($v_{\rm OH}$=$v_{\rm p}$+$v_{\rm e}$) from the half-width of the trapezoid's lower base (Figure \ref{fig:1}), where $v_{\rm p}$ is the parent molecule outflow velocity and $v_{\rm e}$ is the OH ejection velocity \citep{1990AA...238..382B}.
Gaussian-fitting of full width at half maximum (FWHM) offers an alternative velocity measure \citep{1999AJ....118.1850B}, which we applied to derive the Doppler shift relative to the cometary nucleus rest frame (Table~\ref{tab:3}).


The production rates of OH were derived from the integrated flux density, taking into consideration the quenching effect of OH (see detail in e.g. \citealt{1985AJ.....90.1117S,1990AA...238..382B,2023AA...677A.157D,2025AA...701A.204L}).
The population inversions of the $\Lambda$-doublet were primarily controlled by 3I’s heliocentric velocity ($v_{\rm h}$) due to Doppler-shifted solar UV excitation.
All the parameters were listed in Table~\ref{tab:3}.

\begin{deluxetable*}{cccccccccc}
\setlength{\tabcolsep}{3mm}
\tablecaption{Spectral Characteristics of 3I/ALTAS \label{tab:3}}
\tablehead{
\multirow{2}{*}{Species} & \colhead{$\langle r_{h} \rangle$} & \colhead{$ \rm FWHM $$^{\rm a}$} & \colhead{$S^{\rm b}$} & \colhead{Doppler shift$^{\rm c}$} & \multirow{2}{*}{$i$$^{\rm d}$} & \colhead{$T_{\rm bg}$$^{\rm e}$} & \colhead{$v_{\rm OH}$$^{\rm f}$} & \colhead{$Q_{\rm OH}$$^{\rm g}$} \\
\colhead{} & \colhead{$[\rm au]$} & \colhead{$[\rm km \cdot s^{-1}]$} & \colhead{$\rm [Jy \cdot km \cdot s^{-1}]$} & \colhead{$\rm [km \cdot s^{-1}]$} &  & \colhead{[K]} & \colhead{$[\rm km \cdot s^{-1}]$} & \colhead{$[\rm 10^{28}\ s^{-1}]$}
}
\startdata
\multirow{3}{*}{OH} & 2.54 & $-$ & $<0.047$ & $-$ & 0.40 & 3.6 & $-$ & $<0.80$ \\
 & 2.27 & $0.553 \pm 0.115$ & $0.065 \pm 0.023$ & $-$0.066 & 0.42 & 3.5 & $0.42\pm0.04$ & $1.32\pm0.47$ \\
 & 1.96 & $0.877 \pm 0.150$ & $0.076 \pm 0.015$ & $-$0.122 & 0.36 & 3.4 & $0.80\pm0.08$ & $1.89\pm0.37$ \\
\enddata
\tablecomments{$^{\rm a}$The FWHM line width obtained from a Gaussian fit; $^{\rm b}$ The integrated intensity or $3\sigma$ upper limit; $^{\rm c}$ The velocity offset of peak temperature; $^{\rm d}$ Maser inversion from \citet{1981AA....99..320D,1988ApJ...331.1058S}; $^{\rm e}$ The background temperature at 1667 MHz was measured by interpolating the continuum maps at 408 MHz \citep{2015MNRAS.451.4311R,1974AAS...13..359H} following the method mentioned in \citet{1981AA....99..320D}; $^{\rm f}$ The expansion velocity; $^{\rm g}$ Production rate.}
\end{deluxetable*}

\subsection{CO lines analysis}

The production rate and expansion velocity of CO in 3I's were derived from the CO(1-0) spectrum.
Using the density distribution based on the Haser model, we adopt a kinetic temperature ($T_{\rm kin}$) of CO of 35 K (see Appendix \ref{sec:spectrum_fit}) and a production rate of water $\rm Q_{H_2O}=10^{28}\, s^{-1}$ (see section \ref{sec:evo}).
The level population of CO was computed by the non-LTE solver for cometary atmospheres from the Planetary Spectrum Generator (PSG, \citealt{2022fpsg.book.....V}).
Collisions with $\rm H_2O$ molecules and electrons, and the radiative pumping by the solar radiation were take into account with an electron density scaling factor $x_{ne}=0.2$ \citep{2025NatAs...9.1476C,2019A&A...630A..19B,2010A&A...518L.150H}.
According to the escape probability method \citep{1987AA...181..169B, 2007AA...473..303Z}, we fitted the CO line profile, the expansion velocity and the production rate were derived (Table~\ref{tab:4}). 
The line profile showed a clear redward offset of the line center, similar to the HCN detection of ALMA in mid-September \citep{2025arXiv251120845R}.


\begin{deluxetable*}{cccccccc}
\setlength{\tabcolsep}{3mm}
\tablecaption{Spectral Characteristics of 3I/ALTAS \label{tab:4}}
\tablehead{
\multirow{2}{*}{Species} & \colhead{$\langle r_{h} \rangle$} & \colhead{$ \rm FWHM $} & \colhead{Doppler shift} & \colhead{$T_{\rm kin}$} & \colhead{$v_{\rm exp}$} & \colhead{$Q_{\rm CO}$} & \colhead{$Q_{\rm CO}/Q_{\rm H_2O}$} \\
\colhead{} & \colhead{$[\rm au]$} & \colhead{$[\rm km \cdot s^{-1}]$} & \colhead{$\rm [km \cdot s^{-1}]$} & \colhead{K} & \colhead{$[\rm km \cdot s^{-1}]$} & \colhead{$[\rm 10^{27}\ s^{-1}]$} & \colhead{[\%]}
}
\startdata
\multirow{3}{*}{$^{12}$CO} & 3.03-2.56 & $-$ & $-$ & 35 & $(0.30)$ & $<3.49$ & $-$ \\
 & 2.33-1.75 & $0.643 \pm 0.242$ & 0.202 & 35 & $0.39\pm 0.14$ & $5.75\pm1.91$ & $28\pm11\%$ \\
 & 3.03-1.75 & $0.537 \pm 0.201$ & 0.166 & 35 & $0.32\pm 0.12$ & $4.05\pm1.37$ & $28\pm15\%$ \\
\enddata
\end{deluxetable*}

\section{Discussion} \label{sec:dis}

\subsection{Pre-perihelion Evolution of $H_2O$ production rate} \label{sec:evo}

\begin{figure}
\centering
    \centering
    \includegraphics[width=0.45\textwidth]{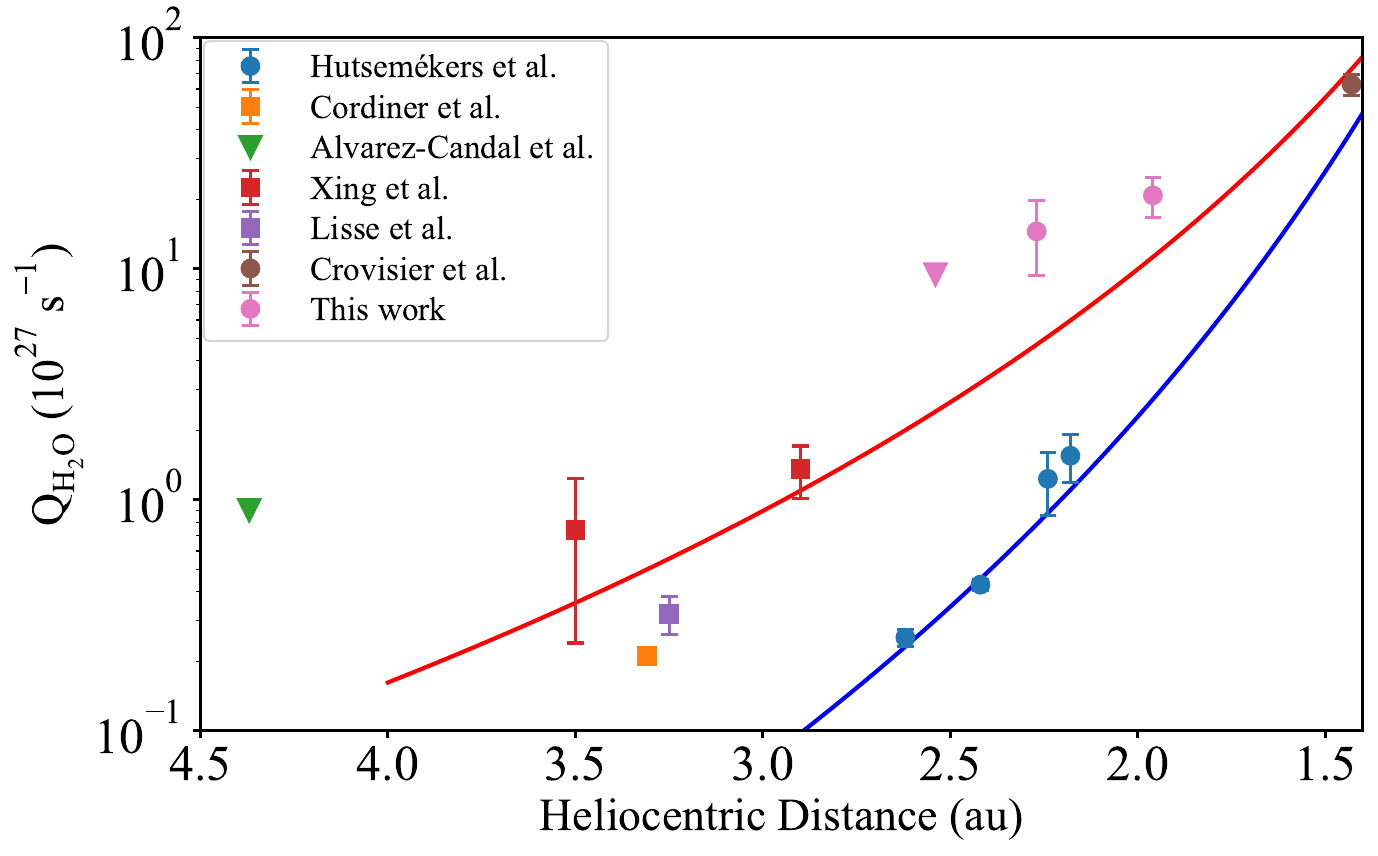}
    \caption{$\rm H_2O$ production rates of comet 3I as a function of heliocentric distance. The measurements from different sources are shown with different colors \citep{2025AA...700L..10A,2025arXiv250926053H,2025ApJ...991L..43C,2025ApJ...991L..50X,2025arXiv251207318L,2025CBET.5625....1C}.
    Error bars represent 1$\sigma$ uncertainties. The symbol of circle, rectangle and triangle indicate the detection via OH, $\rm H_2O$ and upper limits, respectively. The red line shows the best-fit power-law curve with all data excluding those from VLT \citep{2025arXiv250926053H} and JWST \citep{2025ApJ...991L..43C}. The blue one shows the best-fit power-law curve with VLT measurements only.} 
    
    \label{fig:3}
\end{figure}

Figure~\ref{fig:3} shows 3I's pre-perihelion $\rm H_2O$ production rate as a function of heliocentric distance measured with different facilities \citep{2025AA...700L..10A,2025arXiv250926053H,2025ApJ...991L..43C,2025ApJ...991L..50X,2025arXiv251207318L,2025CBET.5625....1C}.
Considering the divergence between different apertures, we separate the water measurements into two datasets. 
By fitting measurements obtained with large apertures ($\gtrsim$ 20000 km, which covers the entire $\rm H_2O/OH$ coma according to the radial surface brightness profiles from \citealt{2025ApJ...991L..50X}), we found a power-law relation of $Q_{\rm H_2O}=(614.55 \pm 101.14)\,r_{\rm h}^{(-5.91 \pm 0.23)} \times 10^{27}\rm \,s^{-1}$ (solid red line in Figure \ref{fig:3}).
In comparison, we also fit the VLT measurements with a much smaller aperture of 3000--4000 km \citep{2025arXiv250926053H} and derived a power-law relationship of $Q_{\rm H_2O}=(825.95 \pm 747.57)\,r_{\rm h}^{(-8.49 \pm 1.01)} \times 10^{27}\rm \,s^{-1}$ (solid blue line) in Figure \ref{fig:3}.
OH measurements were converted to $\rm H_2O$ by multiplying 1.1 \citep{1989AA...213..459C}.
In general, both slopes are consistent with the coma brightness slope and methanol evolution (from -5 to -8; \citealt{2025arXiv251120845R,2025arXiv251025035Z}).
As mentioned in the \citet{2025arXiv251025035Z}, the growth of brightness before $\sim2$ au is much steeper than earlier trend \citep{2025ApJ...990L...2J}, which may result from intensified $\rm H_2O$ sublimation after 3I crossed the $\rm H_2O$ snowline (Figure~\ref{fig:3}; also see Figure 2 in \citealt{2025arXiv251119112T}). 

\subsection{An effective extended-source interpretation}
\label{secondary sub}

The discrepancy between $\rm H_2O$ production rate measured with different sizes of apertures is obvious, as already discussed in several previous works on 3I's activity both pre-perihelion \citep{2025ApJ...991L..43C,2025ApJ...991L..50X,2026arXiv260115443T} and post-perihelion \citep{2025arXiv251222354C}.
Moreover, water ice was detected in 3I's coma at an early stage beyond 4 au \citep{2025ApJ...992L...9Y}.
These evidences point to possible contribution from water sublimation from icy particles in the coma.
To quantify the contribution of extended source of water, we use an aperture-dependent two-component parameterization model of the  production rate, and calculate the extended source fraction.

\begin{figure*}[htbp]
	\centering
	\subfigure[]{
		\centering
		\includegraphics[width=0.45\linewidth]{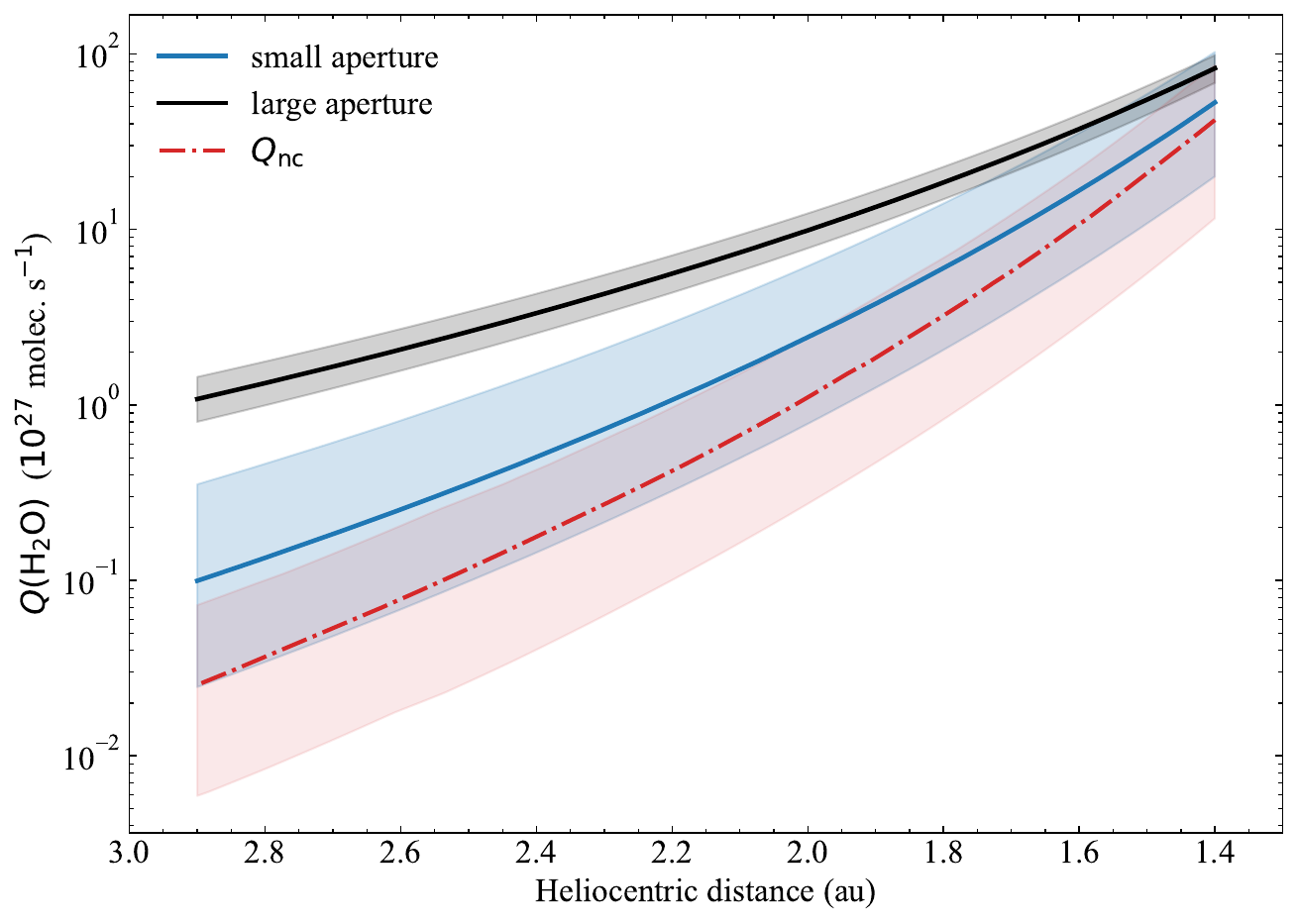}
		\label{4:a}}
	\subfigure[]{
		\centering
		\includegraphics[width=0.45\linewidth]{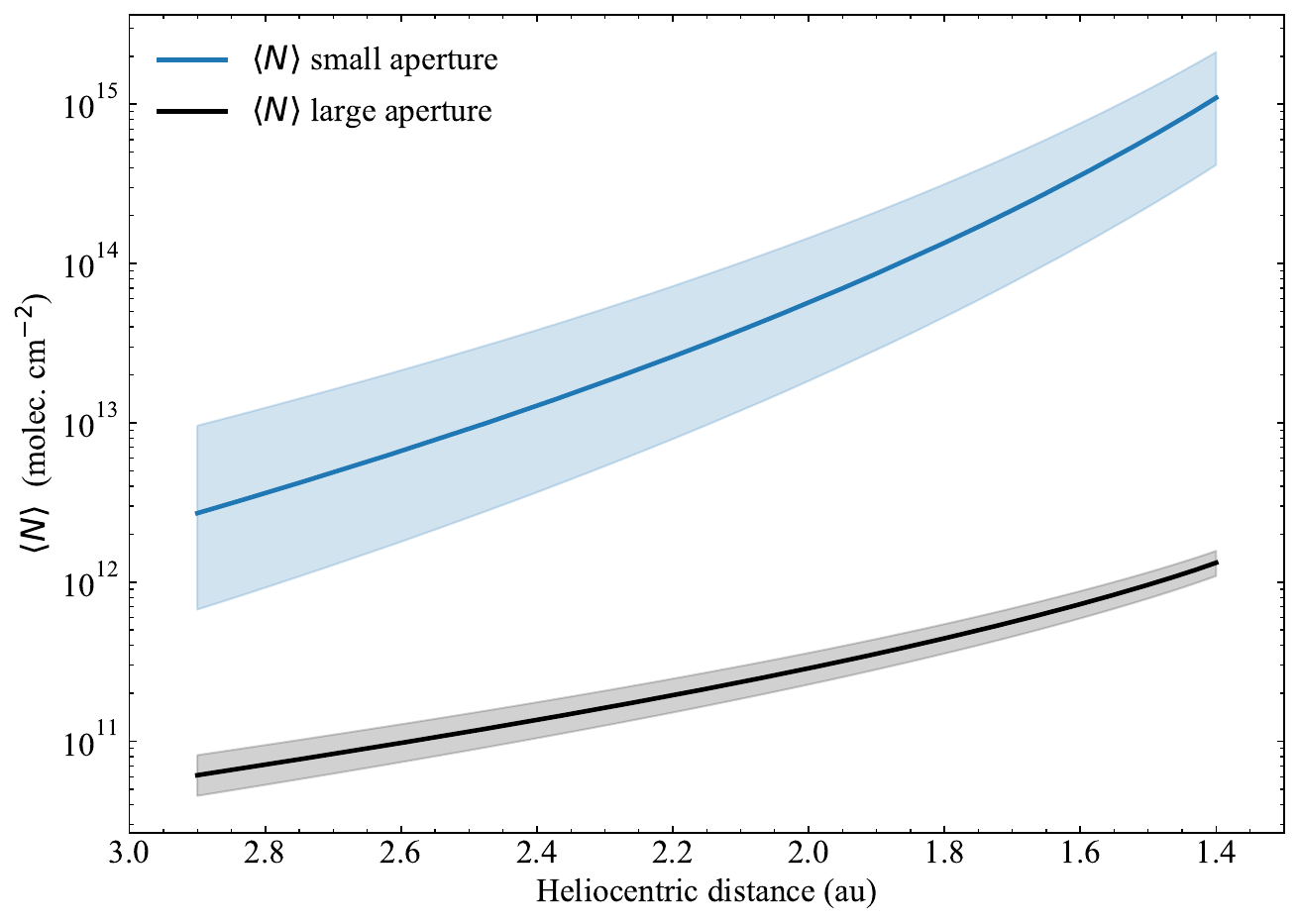}
		\label{4:b}}
    \\	
	\subfigure[]{
		\centering
		\includegraphics[width=0.45\linewidth]{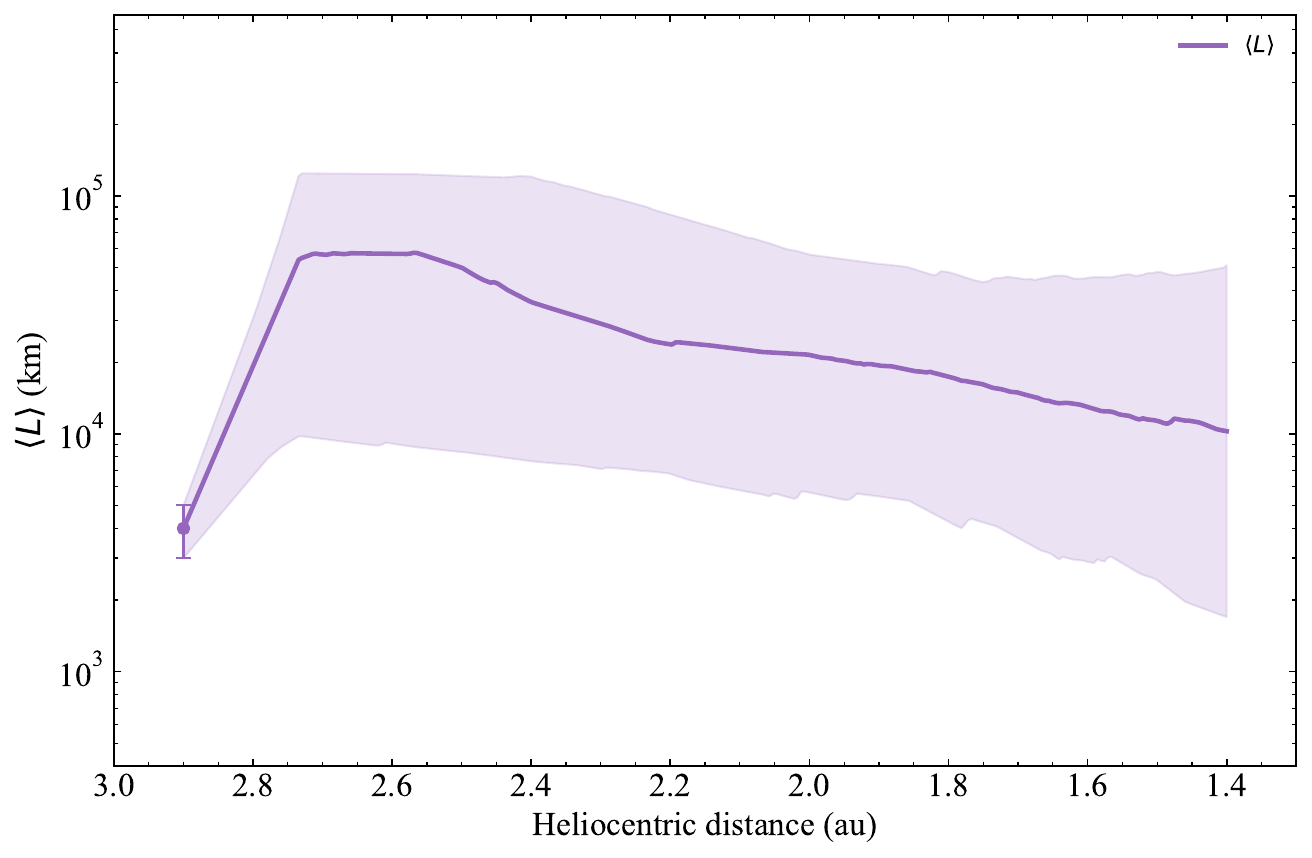}
		\label{4:c}}
	\subfigure[]{
		\centering
		\includegraphics[width=0.45\linewidth]{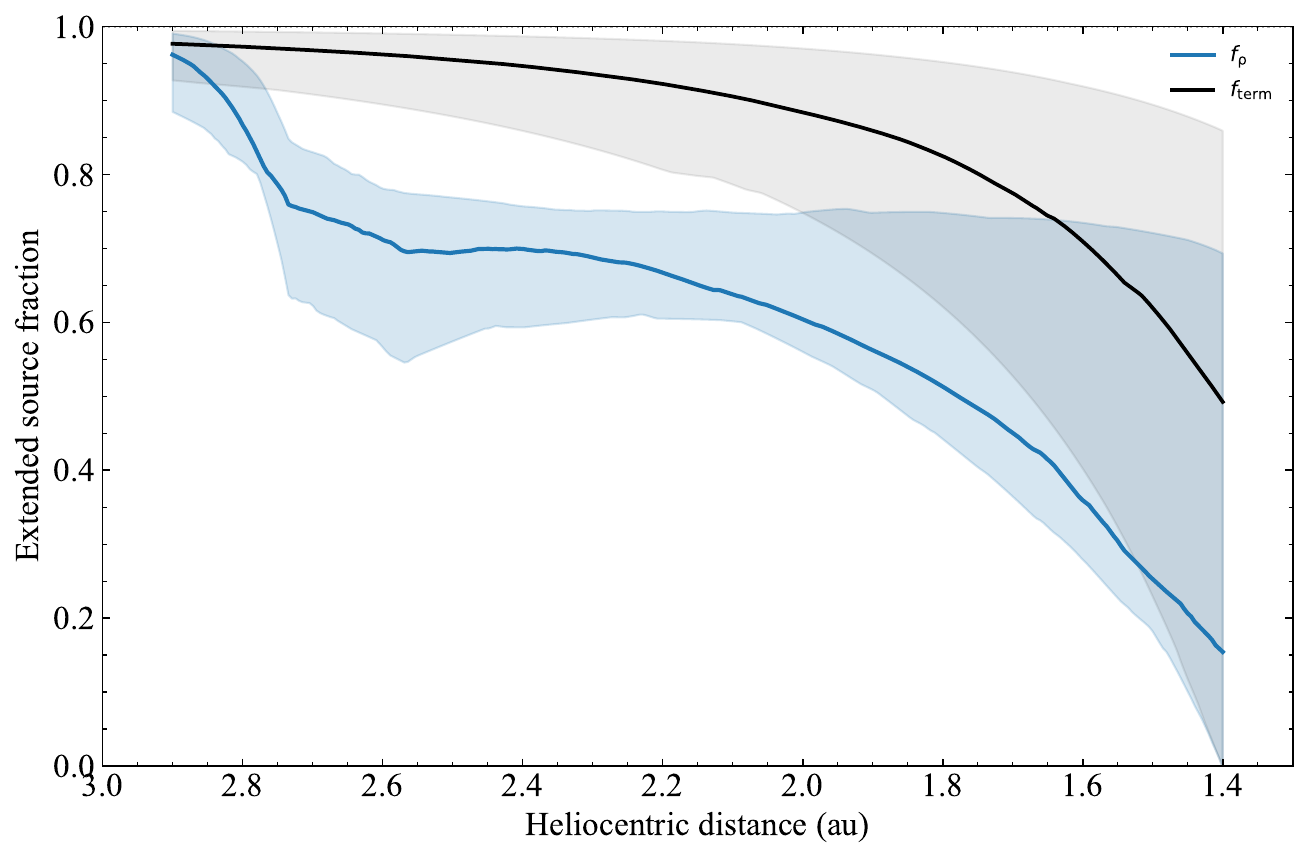}
		\label{4:d}}
	
    \caption{Contribution of extended source. (a) The power-law relationship between $\rm H_2O$ production rate and heliocentric distance. Black and blue solid lines are the fitted power-law relations with measurements done with large apertures ($\gtrsim20000$ km) and small apertures (3000--4000 km), respectively. Red dashed line is the curve of production rate directly from the nucleus inferred from the model fitting. (b) Aperture-dependent column density calculated based on the production rate shown in (a). (c) Best-fitted equivalent scale-length of $\rm H_2O$ extended source with an initial of $L_{\rm ext}=4000^{+1000}_{-1000}$ km at 2.9 au is derived from \citet{2025ApJ...991L..50X}. (d) Fraction of the extended source in the total water production within different apertures varying against heliocentric distances.}
    \label{model results}
\end{figure*}

For a circular aperture of projected radius $\rho$ (km), the effective production rate at heliocentric distance $r_h$ is parameterized as:
\begin{equation}
\label{equ:ex}
    Q_\rho=Q_{\rm nc}+(Q_{\rm term}-Q_{\rm nc})[1-{\rm exp}({-\rho/L_{\rm ext}})]\,.
\end{equation}
We employed the term $L_{\rm ext}(r_h)$ as the scale length of extended source at heliocentric distance $r_h$, defined as the nucleocentric distance where the contribution of extended source in aperture $\rho$ reaches ($1-1/e$) of its total contribution (the difference between the nuclear production rate $Q_{\rm nc}$ and the terminal production rate $Q_{\rm term}$), and shows how rapidly the extended-source contribution saturates with increasing aperture.
We consider the observations carried out with VLT/UVES representative of of ``small-aperture'' measurements \citep{2025arXiv250926053H} which has an equivalent circle aperture $\rho_{\rm eq}$ is $1.93''$ (Appendix \ref{A}), corresponding to $\rho=$[3360-3920] km at the distance of [2.4-2.7] au.
Measurements reported by observations with facilities of large apertures, namely Swift (aperture size scaled at $\rho=10''$, \citealt{2025ApJ...991L..50X}), SPHEREx ($\rho=12''$, \citealt{2025arXiv251207318L}), Nan\c{c}ay ($\rho_{\rm eq}=3.77'$, \citealt{2025CBET.5625....1C}) and TMRT ($\rho=5.8'$, this work), are considered to represent 3I’s total water production rate consisting both the nucleus emission and contribution from the extended source.
We used a statistical model based on Monte-Carlo (MC) method to estimate the amount of $\rm H_2O$ produced via direct nucleus sublimation vs sublimation from icy grains in the extended coma (see details in Appendix \ref{A}).

We found the best-fit $\rm H_2O$ production rate directly released from the nucleus to be $Q_{\rm nc}=(1140_{-783}^{+980})r_h^{(10.30_{-1.09}^{+0.20})}\times 10^{27}\,\rm s^{-1}$ (Figure \ref{4:a}).
Figure \ref{4:b} is a diagnostic conversion from the fitted/assumed $Q(\rm H_2O)$ curves to an average column density within each aperture, using simplified parent-only Haser approximations \citep{2004come.book..391B}, and an expansion velocity of $0.85r_h^{-0.5} \,\rm km \cdot s^{-1}$.
For the fitting of $L_{\rm ext}$, we used the radial brightness profile at 2.9 au observed by Swift \citep{2025ApJ...991L..50X} as the starting point and treated $L_{\rm ext}$ as a free parameter. 
Figure \ref{4:c} indicates $L_{\rm ext}$ experienced a brief period of increase at about 2.9 au to 2.6 au, which is consistent to the increase of radial surface brightness profiles from 3.5 au to 2.9 au in \citet{2025ApJ...991L..50X}.
Due to the increase of the temperature, $L_{\rm ext}$ subsequently decreases slowly when the comet approaches the Sun and the water ice sublimates more rapidly.
We defined the fraction of extended-source in the entire coma 
\begin{equation}
\label{ft}
    f_{\rm term}=\frac{Q_{\rm term}-Q_{\rm nc}}{Q_{\rm term}}\, 
\end{equation}
and in aperture $\rho$
\begin{equation}
\label{fr}
    f_{\rho}=\frac{Q_{\rho}-Q_{\rm nc}}{Q_{\rho}}\,.
\end{equation}

Figure \ref{4:d} shows $f_{\rm term} \sim0.5$ near 1.4 au and $\sim0.9$ at 2.9 au implying that a large fraction of the ``large-aperture'' production is attributed to the extended source.
Note that the fraction may be overestimated outside the 2.6 au as the ``small-aperture'' measurements were all obtained within 2.6 au.
The wide gray band at the large aperture demonstrate that with measurements at two apertures, $Q_{\rm nc}$ and $L_{\rm ext}$ are not sufficient to be confined within a narrow margin of error among MC realizations.
Therefore, the existence of a strong extended contribution is robust, but its exact fraction is only loosely constrained.

The significant contribution of sublimation of extended source in the coma of 3I beyond 2 au from the Sun is similar to what was found with the long-period comet C/2009 P1 (Garradd), whose extended-source (icy grains/chunks) persists at distances beyond $10^6$ km from the nucleus and contributes more than 60\% of the water production rate at $r_h\sim2$ au \citep{2013Icar..225..740C}. 
For other comets with confirmed extended-source sublimation in the coma, their derived ratios also vary across different methods (e.g., 103P at around its perihelion, 25-40\% in \citealt{2025PSJ.....6...95F}, 85\% in \citealt{2025Icar..43516557S}, 77\% in \citealt{2013Icar..225..688F}), and more depending on the observation data used at different $r_h$ (e.g., 46P, \citealt{2021PSJ.....2...55R,2021PSJ.....2..104K,2021PSJ.....2..176P,2021PSJ.....2...45B,2023PSJ.....4...85K}).
Meanwhile, the factors related to the occurrence of extended sources and their contributions remain difficult to constrain.
In most cases, extended sublimation is associated with the activity of highly volatile species, such as $\rm CO_2$ in C/2013 US10 \citep{2018ApJ...862L..16P}, 103P \citep{2014Icar..238..191P}, CO in C/2009 P1 \citep{2014AJ....147...24F}, where these volatiles could eject dust particles with high water ice content, especially at larger heliocentric distances. 
Given the relatively high amount of carbon dioxide measured around 3I \citep{2025ApJ...991L..43C,2025arXiv251207318L}, it is possible that such mechanism is at work for 3I’s secondary sublimation. 
Further confirmation will require high-resolution observations or space missions.

\subsection{CO abundance} \label{sec:abu}

\begin{figure*}
\centering
    \centering
    \includegraphics[width=0.9\textwidth]{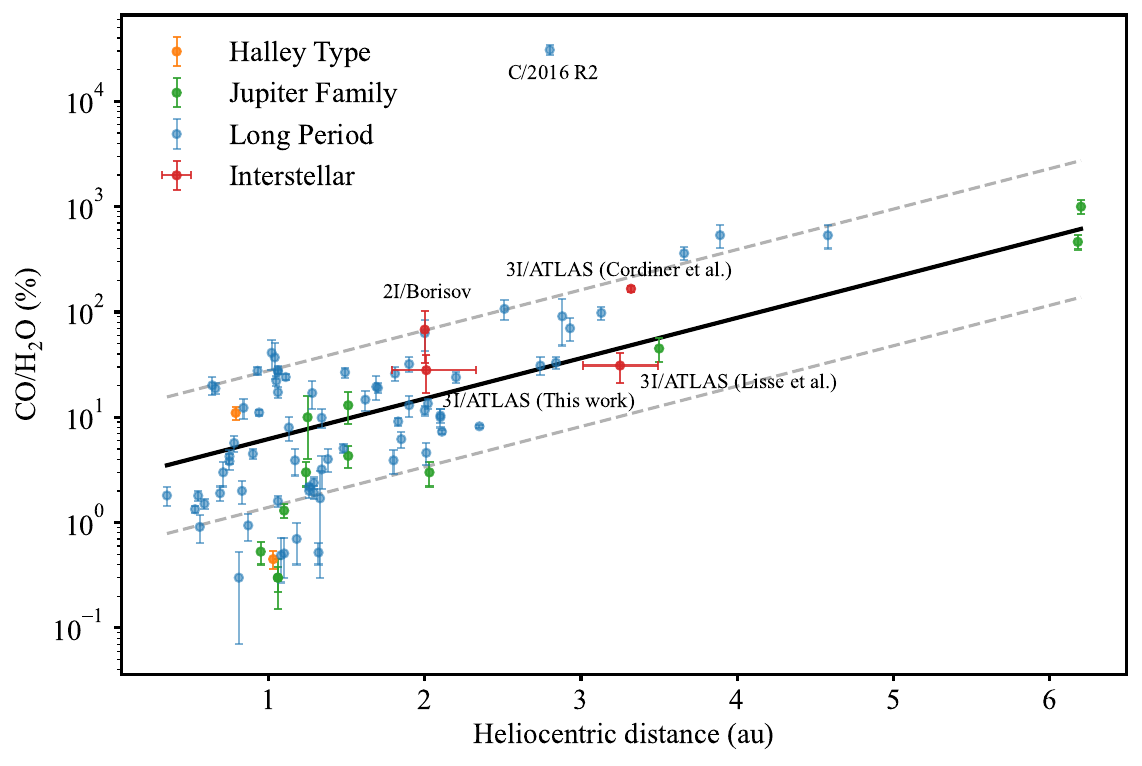}
    \caption{Coma CO/$\rm H_2O$ mixing ratios as a function of heliocentric distance for Solar System comets and interstellar comets. The measurements of different dynamic group are shown with different colors. All data are summarized from the collections in previous works \citep{2016Icar..278..301D,2019AJ....158..128M,2022PSJ.....3..247H} and recent observations \citep{2018A&A...619A.127B,2024A&A...690A.271B,2022A&A...664A..95B,2022PSJ.....3....6D,2025arXiv251105662R,2020NatAs...4..861C,2025PSJ.....6..139W}.A log-linear curve is fitted to the Solar System comets, with $1\sigma$ prediction band shown as dashed gray lines. The other two results of 3I/ATLAS are from \citet{2025ApJ...991L..43C} and \citet{2025arXiv251207318L}.}
    \label{fig:5}
\end{figure*}

3I exhibited a $\rm CO_2$-domainated coma with relative low CO and $\rm H_2O$ occupied in coma during the early stage of its activity beyond 3 au, which results in the very different CO/$\rm H_2O$ ratio about 31\% and 166\% in two works, respectively \citep{2025arXiv251207318L,2025ApJ...991L..43C}. 
Using CO and $\rm H_2O$ production rates retrieved by radio observations, we derived a CO/$\rm H_2O$ ratio of $(28 \pm 11)\%$, which place 3I well above the typical Solar System cometary CO abundance of $\sim4\%$ \citep{2014ApJ...791..122P,2024come.book..459B}.
The value is similar to some “CO-rich” Oort Cloud comets at about 2 au like C/2012 X1 (LINEAR) of $(32\pm5)\%$, C/2009 P1 (Garradd) of $(19\pm2)\%$, C/2008 Q3 (Garradd) of $(26\pm4)\%$ \citep{2012LPICo1667.6330B,2014acm..conf...46B,2020NatAs...4..861C,2022PSJ.....3..247H}, whereas the interstellar comet 2I/Borisov reaches above 60\% \citep{2020NatAs...4..861C,2020NatAs...4..867B} (Figure \ref{fig:5}). 
Within uncertainties, 3I’s CO/$\rm H_2O$ ratio is intermediate between the upper envelope of Solar System comets and the elevated CO abundance reported for 2I/Borisov, suggesting a comparatively CO-enriched volatile inventory relative to canonical Solar System comet compositions.
But $\rm H_2O$ production rate increased more rapidly than CO when 3I got close to the Sun, leading the CO/$\rm H_2O$ ratio decreasing to the level of Oort Cloud comets (Figure \ref{fig:5}).

Comet 3I has an estimated CO/HCN about $230\pm 76$ under an assumption of $Q_{\rm HCN}=2.5\times10^{25}\,\rm s^{-1}$ \citep{2025arXiv251120845R,2025arXiv251002817C}, which is higher than the CO/HCN values reported for most Solar System comets at around 2 au (see Figure 4 in \citealt{2020NatAs...4..861C}), but remains below the unusually large ratio inferred for the interstellar comet 2I/Borisov ($630_{-340}^{+200}$) \citep{2020NatAs...4..861C}.
While 2I/Borisov’s CO/HCN is described as exceeding that of any previously observed comet except the extreme case of C/2016 R2 (PANSTARRS) (CO/HCN=26,400 at $r_h$=2.8 au, \citealt{2018A&A...619A.127B}), 3I’s intermediate value nonetheless also indicates a comparatively CO-enriched composition relative to HCN but more similar to Solar System cometary abundances than 2I.

\section{Conclusions} \label{sec:con}

We conducted radio observations of 3I in August and September, 2025, focusing on the monitoring of OH as a proxy for $\rm H_2O$ and on the detection of CO together with its isotopologues.

From the 18-cm OH maser lines observed with TMRT, we derived OH production rates for two epochs: $(1.32\pm0.47)\times10^{28}\ \rm s^{-1}$ at 2.27 au and $(1.89\pm0.37)\times10^{28}\ \rm s^{-1}$ at 1.96 au, adopting an expansion velocity of $\sim 0.4$--$0.8\ \rm km\,s^{-1}$ for September. 
Combined with results from previous work, we obtained a pre-perihelion heliocentric-distance dependence for $\rm H_2O$ production with a power-law index of $\sim 5$--6, while noting that the rapid rise in water sublimation inside the $\sim 2.7$ au snow line should be taken into account.
Overall, we find that the $\rm H_2O$ production rate increases more rapidly inside the snow line, consistent in brightness with the expected behavior, although the evolution between $\sim 2$ and 3 au before perihelion remains complex and requires additional measurements. 
We note that, MeerKAT obtained near perihelion OH production rate measurements with about $2'$ beam size could provide further constraints on production rate and extended-source fraction \citep{2025ATel17473....1P,2025ATel17499....1P}.

Using the CO($J$=1--0) line, we derived production rates of ($\rm 5.75\pm1.91 \times 10^{27}\, s^{-1}$) at around 2.01 au. 
The CO production rate increased as 3I moved closer to the Sun, but more slowly than $\rm H_2O$. Based on these measurements, we estimate $\rm CO/H_2O=(28\pm11)\%$ at $\sim 2.01$ au. 
This value is lower than the last interstellar comet 2I, but still higher compared to the mean value of Solar System comets.

Based on the $\rm H_2O$ production rates obtained from different apertures, we estimated the fraction of extended sources in the coma of 3I during its pre-perihelion passage. 
The extended-source fraction accounted for approximately over 80\% around 2--3 au, then decreased to about 50\% at perihelion.

\begin{acknowledgments}
The authors thank all the staff of the Tianma-65m Radio Telescope at Shanghai Astronomical Observatory and the 13.7-m Millimeter Wave Telescope at Purple Mountain Observatory for their assistance. This work is financially supported by the National Natural Science Foundation of China (No. 12233003).

\end{acknowledgments}

\facilities{PMO: 13.7m, SHAO: TMRT}

\software{GILDAS and CLASS \citep{2018ssdd.confE..11P}, astropy \citep{2013AA...558A..33A,2018AJ....156..123A,2022ApJ...935..167A}, Planetary Spectrum Generator (PSG; \citealt{2022fpsg.book.....V}), Matplotlib \citep{2007CSE.....9...90H}, Numpy \citep{2020Natur.585..357H}.}


\appendix

\restartappendixnumbering

\section{Spectral modeling}
\label{sec:spectrum_fit}

\subsection{Spectral fitting}

We modeled the observed spectrum by using a spherically symmetric coma \citep{2020PSJ.....1...83H,1957BSRSL..43..740H} and fitting for two global parameters: the expansion velocity $v_{\rm exp}$ and the production rate $Q$. 
The non-LTE excitation calculation provides the radial fractional populations at high and low level ($n_u$ and $n_l$). 
For each trial $(v_{\rm exp},Q)$, we compute a beam-averaged main-beam brightness temperature spectrum $T_{\rm mb}(v)$ by integrating the line emissivity along the line of sight and convolving with a 2D Gaussian telescope beam.  
Optical-depth effects are accounted for using an escape-probability factor $\beta(\tau)$ in an LVG/Sobolev approximation (\citealt{1984mrt..book...21R}, see details in \citealt{1997PhDT........51B} and \citealt{1987AA...181..169B}).
The microscopic line profile is taken as a Gaussian, while the bulk outflow is mapped into the spectrum through the projected line-of-sight velocity field.
The model spectrum is computed by integrating the emissivity along the line of sight and averaging over a 2D Gaussian telescope beam. 
The final beam-averaged specific intensity is normalized by the numerical beam integral.
The fit is performed over a defined velocity window around line center, while the final best-fitting model is evaluated and plotted over the full velocity range for presentation. 
Parameter uncertainties are estimated from Monte-Carlo resampling of the spectrum using the measured channel RMS.
We report conservative $1\sigma$ errors based on these estimates.

\subsection{Assumption of Kinetic Temperature}

In considering the kinetic temperature of the coma, we adopted a range of 20–50 K, consistent with values used in \citet{2020NatAs...4..861C} and \citet{2025arXiv251002817C} at approximately 2 au. This range is also supported by the radial temperature profile derived from $\rm CH_3OH$ lines on September 22 at large nucleocentric distances.
In the case of $x_{ne} = 0.2$, the derived CO production rates were $(4.60 \pm 1.58) \times 10^{27}~\rm s^{-1}$ at 20 K and $(6.81 \pm 2.24) \times 10^{27}~\rm s^{-1}$ at 50 K.
Since the choice of kinetic temperature within this range does not significantly affect the production rate, we adopted a mean value of 35 K for further analysis in the main text.

\section{Water production rate with extended source}
\label{A}

We model the aperture dependence of the retrieved $\rm H_2O$ production rate with a two-component prescription: a nucleus source plus a extended source (inferred from the relationship between column density and nucleocenteric distance used in \citet{2014Icar..238..191P}, similar to the trend in \citealt{2025PSJ.....6...95F} and the Q-curve in \citealt{1996ApJ...464..457X,1998Icar..135..377D,2001Icar..153..361D,2016ApJ...820...34D,2006ApJ...653..774B,2017AJ....153..241B,2021PSJ.....2...45B,2011Icar..216..227V}). 

To compare production rates measured with apertures of different shapes, we convert a rectangular spectroscopic slit to an equivalent circular aperture $\rho_{\rm eq}$. 
In the isotropic, constant-velocity outflow approximation, the column density scales as $N\propto \frac{1}{\rho}$ where $\rho$ is the projected nucleocentric distance. The quantity that enters the aperture-averaged measurement is therefore $I=\int_A \frac{1}{\rho}\,\rm da$, so that $\rho_{\rm eq}=\frac{1}{2\pi}\int_A \frac{1}{\rho}\,\rm da$.
If the slit is a rectangle of length $L$ and width $W$ (e.g., VLT, \citealt{2021Natur.593..372M}), centered on the nucleus, with $a=L/2$, $b=W/2$ and $r=\sqrt{a^2+b^2}$, then
\begin{equation}
    \rho_{\rm eq}=\frac{2}{\pi}[a\ln (\frac{b+r}{a})+b\ln (\frac{a+r}{b})]\,.
\end{equation}

We propagated the uncertainties of the empirical aperture-dependent production-rate laws as stated in Equation \ref{equ:ex} using a Monte Carlo (MC) resampling scheme \citep{jcgm:2008:PDMC,2010arXiv1009.2755A}. 
Each aperture is described by a power law $Q_{\rm nc}=Q_{\rm n_0}r_h^{-n}$ (units: $10^{27}\,\rm s^{-1}$), where $Q_{\rm n_0}$ is the production rate at $r_h=1$ au, $n$ is the power-law index, they are taken with their quoted $1\sigma$ uncertainties. 
In the MC, we draw $n_n$ from a normal distribution $n\sim N(n_0,\sigma_n)$.
We draw $Q_{\rm n_0}$ from $Q_{\rm n_0}\sim N(Q_0,\sigma_Q)$ and enforce $Q_{\rm n_0}>0$ via simple rejection (a few resampling attempts) followed by clipping at zero. 
We assume $Q_{\rm n_0}$ and $n$ are independent and do not include covariance between parameters or between the two apertures. 
For each MC realization we compute the corresponding curves $Q_{\rm small}(r_h)$ and $Q_{\rm large}(r_h)$. At each $r_h$, we summarize the ensemble by the median (50\%) as the central estimate and the 16\%-84\% percentile interval as an uncertainty band $1\sigma$, which corresponds to the central 68\% interval of the MC distribution.
To propagate these observational uncertainties into the secondary-sublimation (extended source) inversion, we perform MC refits in production-space. 
For each selected MC realization, we treat the large-aperture curve as the coma-integrated production rate, $Q_{\rm term}(r_h)\simeq Q_{\rm large}(r_h)$.
We convert the angular aperture radius to a projected physical radius and fit the global model parameters ($Q_{\rm n_0}$, $n$, $L_{\rm ext}$) in a global heliocentric dependence $Q_{\rm nc}=Q_{\rm n_0}r_h^{-n}$ and $L_{\rm ext}$ as a free parameter at each heliocentric distance.
Each refit yields the derived curves $Q_{\rm nc}(r_h)$, $L_{\rm ext}(r_h)$ as well as the fraction of extended-source in total production rate $f_{\rm term}$ (Equation \ref{ft}) and in aperture-dependent production rate (Equation \ref{fr}).
We report MC medians and $1\sigma$ bands for all fitted parameters and derived quantities. 
In Figure \ref{model results}, the displayed “best” curves correspond to the MC median solution for consistency with the MC uncertainty envelopes. 
This MC procedure captures the uncertainty propagated from the empirical power-law fits, while uncertainties in aperture and additional systematic effects (e.g., instrument-to-instrument differences, model-form uncertainty, or parameter correlations) are not included and should be interpreted as potential additional sources of error.

This analysis intentionally treats the $Q(r_h)$ curves as given and does not attempt to homogenize instrumental or excitation/fluorescence differences among different wavelengths. 
Consequently, the inferred parameters (especially $L_{\rm ext}$) should be interpreted as effective quantities that encode both physical coma extension and any systematic differences inherent to the literature derivations. 
In fact, the reasonable prior on $L_{\rm ext}$ has little effect on the total extended-source fraction ($<5\%$), as we assumed a wide range of [3000-5000] km at 2.9 au.
Despite these limitations, the combined small-aperture vs coma-covering constraints provide a practical quantification of the scale dependence and the degree to which an extended source is required to reconcile the reported activity levels.

\section{observing log and spectrum}

\begin{deluxetable*}{ccccccc}[htb]
\setlength{\tabcolsep}{3mm}
\tablecaption{Observation Log of 3I/ATLAS \label{tab:1}}
\tablehead{
\multirow{2}{*}{Date (2025)} & \multirow{2}{*}{UT Time} & \colhead{$\langle r_{h} \rangle$$^{\rm a}$} & \colhead{$\langle \Delta \rangle$$^{\rm b}$} & \colhead{$\langle \alpha \rangle$$^{\rm c}$} & \colhead{$\langle \psi \rangle$$^{\rm d}$} & \colhead{$ \langle v_{r} \rangle$$^{\rm e}$} \\
\colhead{} & \colhead{} & \colhead{$[\rm au]$} & \colhead{$[\rm au]$} & \colhead{$[^\circ]$} & \colhead{$[^\circ]$} & \colhead{$[\rm km \cdot s^{-1}]$} 
}
\startdata
\multicolumn{7}{c}{18-cm OH lines} \\
\hline
Aug. 26 & 08:54-13:00 & 2.68 & 2.60 & 22.0 & 83.9 & $-$7.05 \\
Sep. 01 & 05:54-12:00 & 2.50 & 2.57 & 22.9 & 74.7 & $-$5.20 \\
Sep. 03 & 05:42-12:00 & 2.44 & 2.57 & 23.0 & 71.5 & $-$4.69 \\
Sep. 08 & 05:48-11:42 & 2.30 & 2.56 & 23.2 & 63.9 & $-$3.70 \\
Sep. 09 & 05:48-11:42 & 2.27 & 2.55 & 23.2 & 62.3 & $-$3.81 \\
Sep. 18 & 05:00-10:42 & 2.02 & 2.54 & 22.0 & 48.8 & $-$3.68 \\
Sep. 23 & 03:42-09:42 & 1.89 & 2.52 & 20.6 & 41.4 & $-$4.40 \\
\hline
\multicolumn{7}{c}{$^{12}$CO($J=1-0$) lines} \\
\hline
Aug. 15 & 10:25-12:59 & 3.03 & 2.66 & 19.1 & 101.6 & $-$13.04 \\ 
Aug. 16 & 10:06-12:57 & 3.00 & 2.65 & 19.4 & 99.8 & $-$12.41 \\ 
Aug. 18 & 09:51-11:37 & 2.93 & 2.64 & 20.0 & 96.7 & $-$11.27 \\ 
Aug. 24 & 09:23-12:27 & 2.75 & 2.60 & 21.6 & 87.1 & $-$8.09 \\ 
Aug. 30 & 08:09-10:26 & 2.56 & 2.58 & 22.6 & 77.8 & $-$5.92 \\ 
Sep. 07 & 08:02-09:19 & 2.33 & 2.56 & 23.2 & 65.4 & $-$4.13 \\ 
Sep. 11 & 07:43-09:04 & 2.21 & 2.55 & 23.0 & 59.3 & $-$3.77 \\
Sep. 13 & 06:56-09:11 & 2.16 & 2.55 & 22.8 & 56.3 & $-$3.68 \\
Sep. 15 & 06:52-08:49 & 2.10 & 2.54 & 22.6 & 53.3 & $-$3.70 \\
Sep. 17 & 06:17-08:43 & 2.05 & 2.54 & 22.2 & 50.3 & $-$3.81 \\
Sep. 23 & 06:03-08:16 & 1.89 & 2.52 & 20.6 & 41.4 & $-$4.51 \\
Sep. 25 & 06:16-08:40 & 1.84 & 2.52 & 19.8 & 38.5 & $-$4.85 \\
Sep. 27 & 05:50-08:06 & 1.79 & 2.51 & 18.9 & 35.5 & $-$5.33 \\
Sep. 29 & 05:26-07:45 & 1.75 & 2.51 & 18.0 & 32.6 & $-$5.85 \\
\enddata
\tablecomments{$^{\rm a}$Mean heliocentric distance during observation time; $^{\rm b}$Mean geocentric distance; $^{\rm c}$Mean solar phase angle (Sun-object-Earth); $^{\rm d}$Mean solar elongation angle (Sun-Earth-object); $^{\rm e}$Mean radial velocity.} 
\end{deluxetable*}

\begin{figure*}[ht]
\plotone{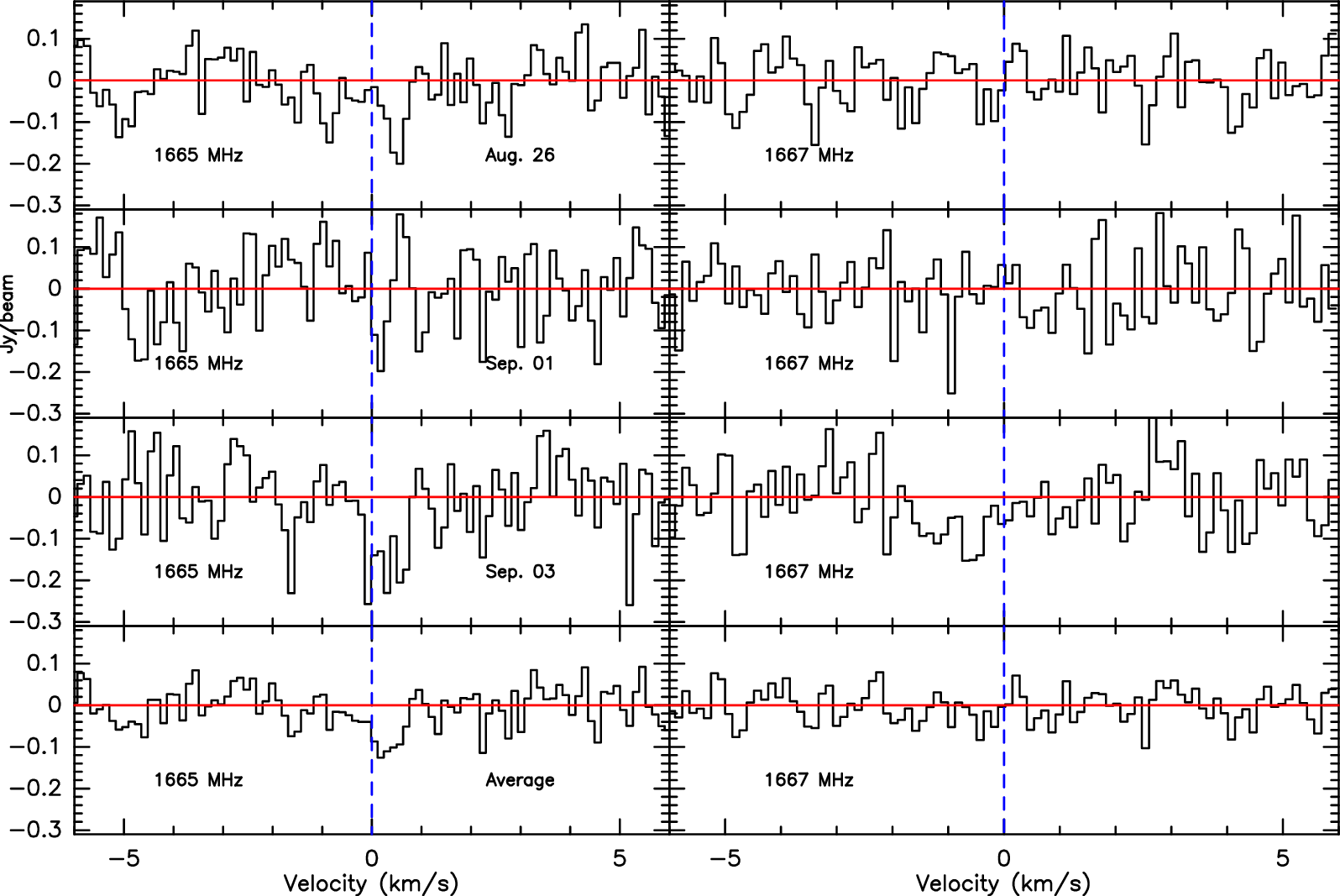}
\caption{Averaged 18 cm OH lines of 3I/ATLAS observed on three separate days: August 26 (top), September 1 (upper middle), September 3 (lower middle), 2025, weighted averages of the 1667 and 1665 MHz lines (bottom). 1665 and 1667 MHz lines are shown in left and right, respectively. The vertical scale is scaled to the flux intensity, and the horizontal scale is the Doppler velocity in the comet rest frame. The velocity resolution is 0.1288 $\rm km \cdot s^{-1}$ after smooth.
\label{fig:all1}}
\end{figure*}

\begin{figure*}[ht]
\plotone{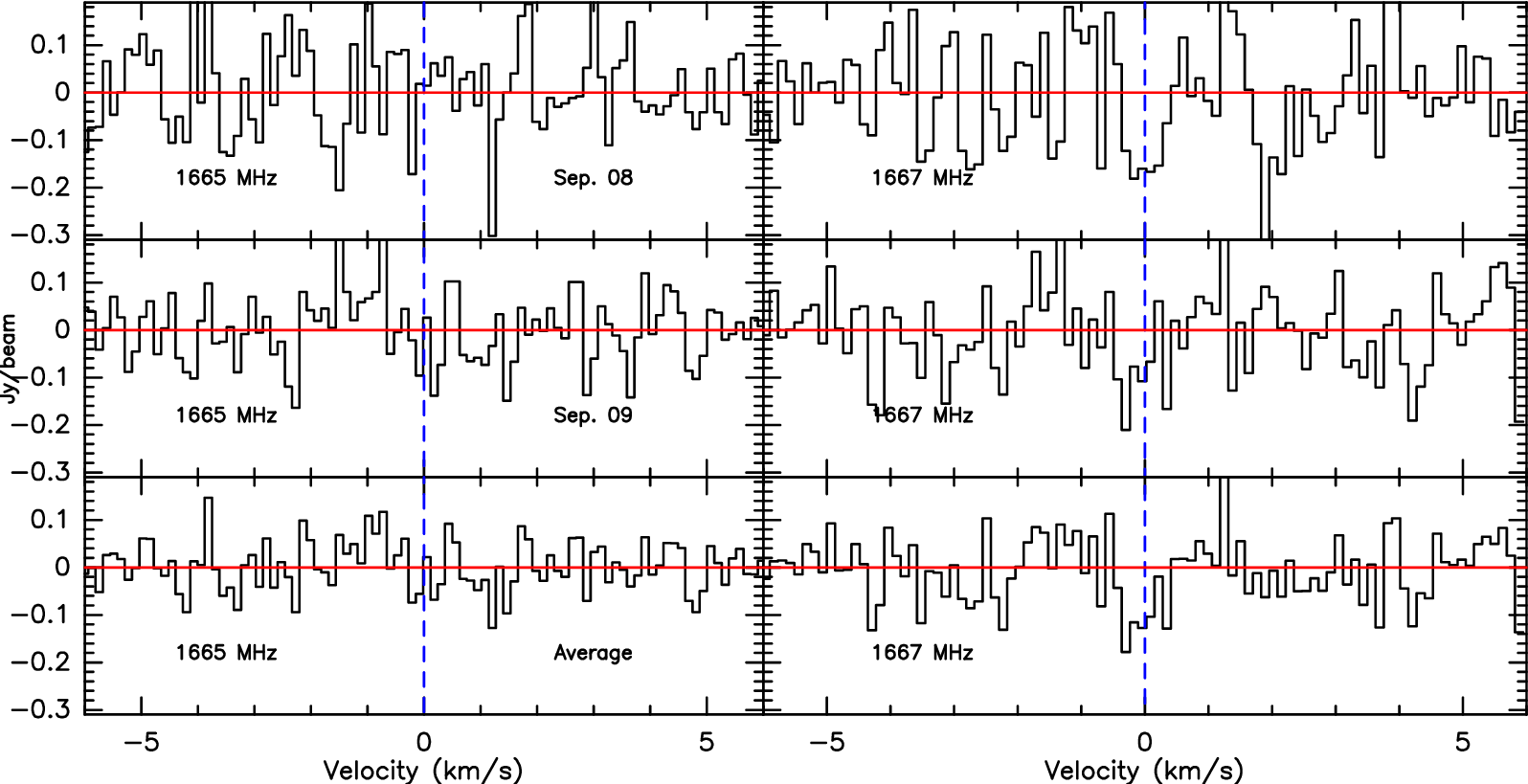}
\caption{Averaged 18 cm OH lines of 3I/ATLAS observed on two separate days: September 8 (top), September 9 (middle), 2025, and weighted averages of the 1667 and 1665 MHz lines (bottom). 
\label{fig:all2}}
\end{figure*}

\begin{figure*}[ht]
\plotone{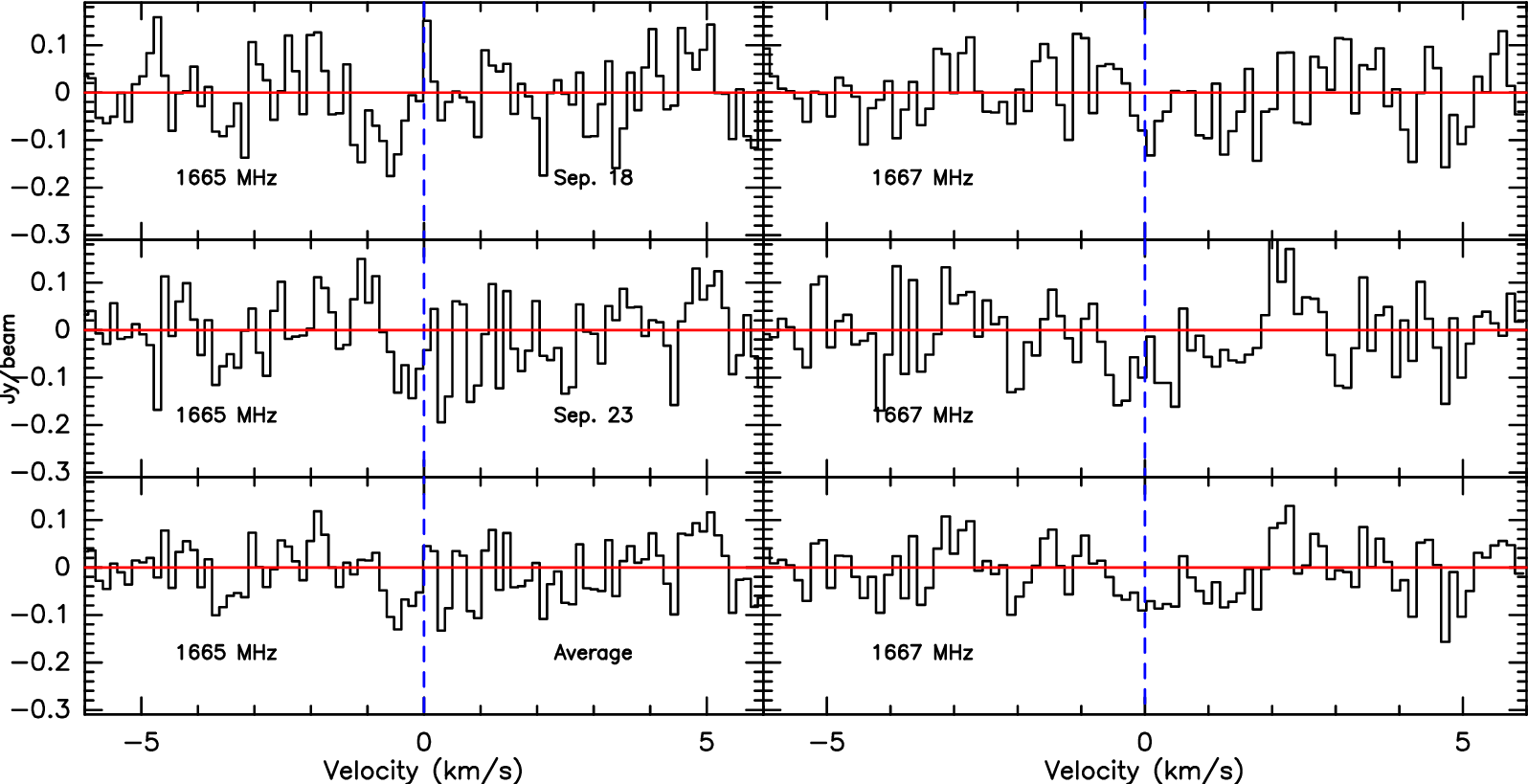}
\caption{Averaged 18 cm OH lines of 3I/ATLAS observed on two separate days: September 18 (top), September 23 (middle), 2025, and weighted averages of the 1667 and 1665 MHz lines (bottom).
\label{fig:all3}}
\end{figure*}

\begin{figure*}[ht]
\begin{center}
    \includegraphics[width=0.6\textwidth]{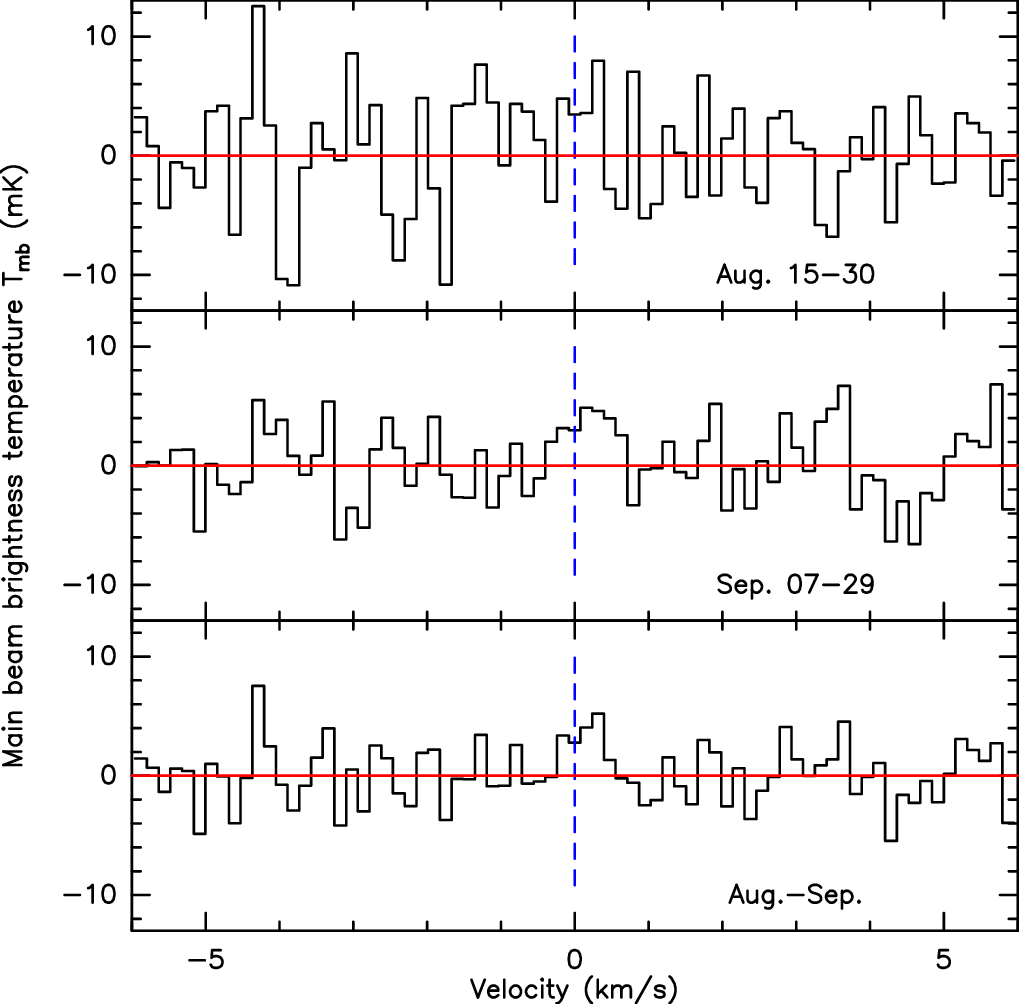}
\end{center}
\caption{Averaged spectra of CO($J$=1-0) in 3I/ATLAS observed in August (top), September (middle), and from August to September (bottom).
The red line is the base line.
The vertical scale is the main beam brightness temperature and the horizontal scale is the Doppler velocity in the comet rest frame.
The velocity resolution is 0.1587 $\rm km \cdot s^{-1}$ after smooth.
\label{fig:allco}}
\end{figure*}

\onecolumngrid 

\section{$\rm H_2O$ and CO production rates}

\begin{deluxetable*}{cccccc}
\setlength{\tabcolsep}{3mm}
\tablecaption{Pre-perihelion $\rm H_2O$ and CO production rate of 3I/ALTAS \label{tab:5}}
\tablehead{
\multirow{2}{*}{Date (2025)} & \colhead{$\langle r_{h} \rangle$} & \colhead{$Q_{\rm H_2O}$} & \colhead{$Q_{\rm CO}$} & \multirow{2}{*}{Instrument} & \multirow{2}{*}{Reference} \\
\colhead{} & \colhead{$[\rm au]$} & \colhead{$[\rm 10^{27}\ s^{-1}]$} & \colhead{$[\rm 10^{27}\ s^{-1}]$} & \colhead{} & \colhead{}
}
\startdata
Jul. 04 & 4.37 & $<0.90$ & $-$ & VLT & \citet{2025AA...700L..10A} \\
Jul. 31-Aug. 01 & 3.50 & $0.74\pm0.50$ & $-$ & Swift & \citet{2025ApJ...991L..50X}\\
Aug. 06 & 3.31 & $0.22\pm0.01$ & $0.37\pm0.02$ & JWST &\citet{2025ApJ...991L..43C}\\
Aug. 01-15 & 3.25 & $0.32\pm0.06$ & $0.1\pm0.03$ & SPHEREx & \citet{2025arXiv251207318L}\\
Aug. 18-20 & 2.90 & $1.36\pm0.05$ & $-$ & Swift &\citet{2025ApJ...991L..50X}\\
Aug. 28 & 2.64 & $0.25\pm0.02$ & $-$ & VLT &\citet{2025arXiv250926053H}\\
Aug. 26-Spt. 03 & 2.54 & $<9.46$ & $-$ & TMRT & This work \\
Sep. 03-04 & 2.44 & $0.43\pm0.02$ & $-$ & VLT &\citet{2025arXiv250926053H}\\
Sep. 07-29 & 2.01 & $-$ & $5.75\pm1.91$ & 13.7m & This work \\
Sep. 08-09 & 2.27 & $14.52\pm5.17$ & $-$ & TMRT & This work\\
Sep. 10 & 2.25 & $1.23\pm0.37$ & $-$ & VLT &\citet{2025arXiv250926053H}\\
Sep. 12 & 2.19 & $1.55\pm0.36$ & $-$ & VLT &\citet{2025arXiv250926053H}\\
Sep. 18-23 & 1.96 & $20.79\pm4.07$ & $-$ & TMRT & This work\\
Oct. 13-16 & 1.43 & $57\pm 6$ & $-$ & NRT & \citet{2025CBET.5625....1C} \\
\enddata
\end{deluxetable*}


\bibliography{sample701}{}
\bibliographystyle{aasjournalv7}



\end{document}